\documentclass{elsart}
\usepackage[dvips]{epsfig}
\begin{document}
\begin{frontmatter}
\title{\nuc{16}{N} as a calibration source
for Super-Kamiokande}

\collab{The Super-Kamiokande Collaboration}

\author[lsu]{E. Blaufuss\thanksref{corresponding}}, 
\author[umd]{G. Guillian},
%ICRR
\author[icrr]{Y. Fukuda},
\author[icrr]{S. Fukuda},
\author[icrr]{M. Ishitsuka}, 
\author[icrr]{Y. Itow},
\author[icrr]{T. Kajita}, 
\author[icrr]{J. Kameda}, 
\author[icrr]{K. Kaneyuki}, 
\author[icrr]{K. Kobayashi}, 
% ...

\address[icrr]{Institute for Cosmic Ray Research, University of Tokyo, Kashiwa,
Chiba 277-8582, Japan}
% ...
\address[lsu]{Department of Physics and Astronomy, Louisiana State 
University, Baton Rouge, LA 70803, USA }
\address[umd]{Department of Physics, University of Maryland, 
College Park, MD 20742, USA }
% ...
\thanks[corresponding]{Corresponding author: University of Maryland,
Department of Physics, College Park, MD 20742, USA,
Tel: +1 301 405 6077, Fax: +1 301 699 9195, e-mail blaufuss@umdgrb.umd.edu}
\thanks[poland]{Supported by the Polish Committee for Scientific Research
Grant 2P03B05316}

\begin{abstract}
The decay of \nuc{16}{N} is used to
cross check the absolute energy scale calibration for 
solar neutrinos established by the electron linear accelerator (LINAC).  
A deuterium-tritium neutron generator
was employed to create \nuc{16}{N} via the (n,p) reaction on \nuc{16}{O}
in the water of the detector.   This technique is isotropic
and has different systematic uncertainties than the LINAC.
The results from this high statistics
data sample agree with the absolute energy scale of the LINAC to better
than 1\%.  A natural
source of \nuc{16}{N} from the capture of $\mu^{-}$ on
\nuc{16}{O}, which is collected as a background to the 
solar neutrino analysis, is also discussed.

\end{abstract}
\end{frontmatter}

\section{Introduction}

The deficit of neutrinos coming from the sun, known as the
solar neutrino problem, has long been established by past
experiments\cite{others1,others2,others3,others4,jnb98}.  
A global analysis of these experimental
results suggests an energy dependent suppression in the flux
of solar neutrinos\cite{Lang}.   This suppression, via neutrino
oscillations\cite{oscpapers}, could produce spectral distortions in the 
measured solar neutrino spectrum.  Super-Kamiokande is the first
experiment in a new generation of neutrino observatories that set out
to measure the spectrum of neutrinos from the sun\cite{skflux,skdn,skspec}.  

Super-Kamiokande (SK), a water Cherenkov detector, is sensitive 
only to the high energy \nuc{8}{B} and rare HEP
neutrinos from the sun.  Detection of lower energy neutrinos
is impeded by the high energy threshold of a water
detector.  The recoil electrons resulting from the
elastic scatter of these neutrinos are observed in SK, providing energy
and directional information in real time.  Since the angular resolution
of the detector for electrons of these energies is limited 
by multiple Coulomb scattering,
performing a kinematic reconstruction of the incident neutrino energy
is precluded.  The measured electron energy is a lower limit 
of the neutrino energy and
the shape of the solar neutrino energy spectrum must
be inferred from the measured recoil electron spectrum.
This situation increases the sensitivity to errors in the absolute energy
scale when making solar neutrino flux and spectrum measurements.
To set the energy scale, an electron linear accelerator (LINAC)
was installed at SK to inject electrons of known energies into
the detector\cite{linac}.  In order to cross check
this calibration, the decay of \nuc{16}{N} is used.
First, a portable neutron generator has been employed
to obtain high statistics data samples of \nuc{16}{N} at
many positions in the detector to accurately check the
absolute energy scale.  Second, the rate of \nuc{16}{N} from the
capture of stopped $\mu^{-}$ is measured as a check of
the solar neutrino analysis tools.

\section{Solar neutrinos in Super-Kamiokande}

Super-Kamiokande is a water Cherenkov detector located in the Kamioka Mine
in Gifu, Japan.  The walls of the 
cylindrical detector are constructed from welded
stainless steel plates, backed with concrete.  The detector is
divided into an inner and and outer detector (ID and OD respectively) by a
stainless steel frame structure that serves as an optical barrier
and a mounting point for all photomultiplier tubes (PMTs).
Cherenkov light in the ID is collected by 11,146 inward facing
50 cm PMTs mounted uniformly on the wall, providing 40\% 
photocathode coverage.  In the OD, 1,885 20 cm PMTs monitor
the 2.5 meter thick veto region.  The veto is 
used to tag incoming particles and serves as
a passive shield for gamma activity 
from the surrounding rock.  The ID encloses
32,500 metric tons of water in a volume that is 36.2 m in height
and 33.8 m in diameter.  The fiducial volume for the solar neutrino
analysis starts 2 m inward of the walls of the ID and contains
22,500 metric tons of water.  The ID is accessible
by a set of calibration ports 30 cm in diameter
leading from the top of the tank directly to the ID.

Solar neutrinos measured in SK have energies that range
from 5 to 18 MeV.  At these energies, the recoil electron
is limited to a few centimeters in range and 
the vertex position is found using the relative timing of
hit PMTs, assuming all Cherenkov photons
came from a single point.  Once a vertex
is reconstructed, the direction of the electron is determined 
using the characteristic shape of the emitted Cherenkov radiation.  
Since roughly 6 photoelectrons in total are collected per MeV of energy, very
few PMTs will have more than one photoelectron
and the number of hit PMTs is used as a measure of the energy.
This number is corrected for PMT dark noise, absorption and scattering
of Cherenkov photons,
and geometrical acceptance of the PMTs, which depends on the 
reconstructed vertex position and
direction in the detector.  This corrected number of hit PMTs must be
properly translated into energy and should be uniform in all 
directions and throughout the volume.

The LINAC calibration provides this translation from the corrected
number of hit PMTs to total energy.  The LINAC\cite{linac} is located
15 meters from the top of SK, inset into the rock wall of the cavern.  
The electron beam is directed in a beam pipe 
through 9 meters of rock shielding, across the top
of the SK tank, and down into the water by a series of bending and
focusing magnets.  The beam momentum is adjustable from 5 to 16 MeV,
matching the range of energies of \nuc{8}{B} solar neutrinos.
The absolute energy of the beam is measured with
a germanium detector.   LINAC data taken at several different
positions and energies are used to set the absolute energy scale
of the Monte Carlo (MC) detector simulation.  The MC extrapolates
this calibration to the entire range of energies, over
the entire volume, in all directions.  The resulting absolute energy
scale is thought to be known with better than 1\% 
uncertainty.  To be confident with such a precise measurement,
a secondary check with a calibration source of comparable precision
is desirable.

The LINAC calibration has limitations.  The electrons are only
moving in a downward direction when they exit 
the beam pipe, possibly introducing 
systematic errors due to direction dependences of the detector.
The presence of the beam pipe in the tank while calibration
data are taken is another limitation.  While this is
modeled in the simulation, it is still the largest source
of systematic error for the calibration, especially at low energies.
Additionally, the beam pipe and equipment associated
with the LINAC calibration can only be operated at a restricted set of
calibration ports, so the calibration must
be extrapolated to the entire fiducial volume.  Operating
the LINAC also requires a great deal of manpower and results in
significant detector down time.

The deuterium-tritium neutron generator (DTG) pulsed calibration source
was built to address the limitations of the LINAC calibration system, 
while providing a cross check of the absolute energy
scale.   The DTG creates \nuc{16}{N} by the (n,p) reaction
on \nuc{16}{O} in the water of the detector.  
The decay of \nuc{16}{N}, with a Q value of 10.4 MeV,
is dominated by an electron with a 4.3 MeV maximum energy coincident
with a 6.1 MeV gamma ray and is well suited to
check the absolute energy scale for the solar neutrino measurement.
With a half life of $7.13$ sec, \nuc{16}{N} is created
\textit{in situ} by lowering the DTG into the detector on a
computer-controlled crane.  After firing, the DTG is withdrawn,
leaving the produced \nuc{16}{N} to decay removed from the presence
of any calibration equipment.  \nuc{16}{N} decays isotropically,
making direction dependence studies on the energy scale possible.
The calibration system is designed to be portable, permitting operation
at virtually any calibration port, and providing a more 
complete mapping of the position dependence 
of the energy scale.  The DTG is also designed to be
easy to operate, requiring less manpower 
and down time for setup and data taking.  The SNO experiment
uses a similar DTG, externally located to their detector,
as a neutron activation source for
calibration purposes\cite{SNO}.

An additional source of \nuc{16}{N} used for calibration
at SK is the capture of cosmic
ray $\mu^{-}$ on \nuc{16}{O}.  These
events are collected as a natural
background in the solar neutrino analysis.
These events occur at a comparable rate to solar neutrinos, and 
comparing the measured event rate to the expected rate serves as
a check of the solar neutrino signal extraction method.  
These events, although limited in statistics, can also be
used to check the absolute energy scale.

The following sections will describe the two methods by
which \nuc{16}{N} is produced for calibration in SK, 
including a description
of the experimental setup used in taking DTG data,
a discussion of \nuc{16}{N} from the capture of $\mu^{-}$, 
as well as information regarding the simulation
of the \nuc{16}{N} beta decay.  The results of
the analysis of both data sets are also presented.

\section{\nuc{16}{N}-Production and Modeling.}

The \nuc{16}{N} used as a calibration source for SK are produced
by two different mechanisms. The first source 
results from the interaction
of 14.2 MeV neutrons in the water
of the detector.  These neutrons are from a 
deuterium-tritium neutron generator (DTG),
and produce a high statistics sample of \nuc{16}{N} at
a set position in the detector.  The second source 
of \nuc{16}{N} is the capture of stopped $\mu^{-}$ 
in the water of the detector.   These events, like 
solar neutrinos, are uniformly distributed
throughout the detector volume.
For both of these data sets, the beta decay of \nuc{16}{N} is carefully
modeled, and the corresponding MC is compared to the data.
Since the energy scale of the MC
is set by the LINAC calibration, a comparison of \nuc{16}{N}
data to MC serves as a cross check of the energy scale.

\subsection{\nuc{16}{N} from the DTG neutron generator}

\begin{figure}
\begin{center}
\epsfig{file=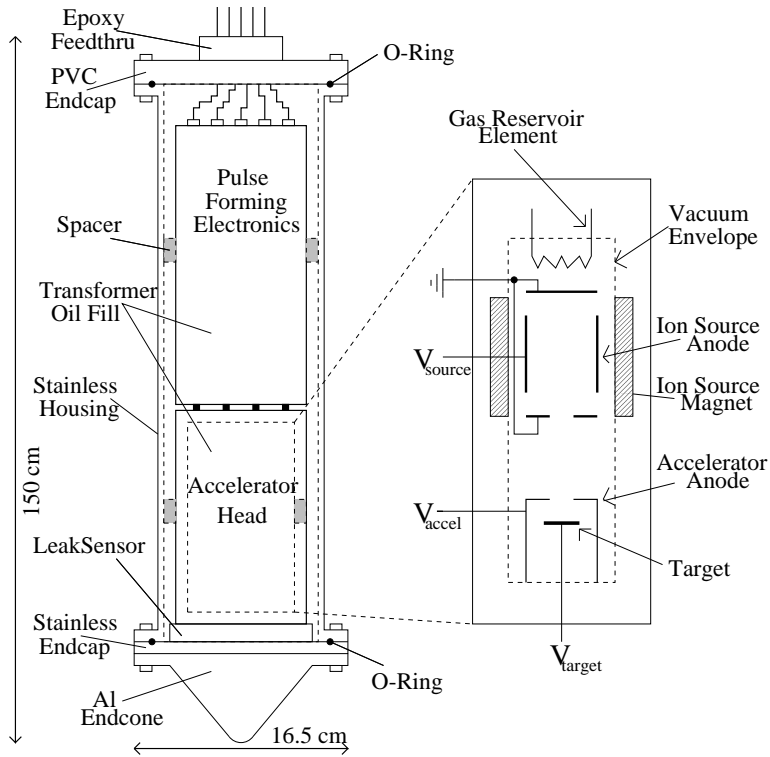,height=10cm}
\end{center}
\caption{Schematic of the DTG setup including 
the stainless steel water-tight housing,
accelerator pulse-forming electronics and accelerator head.
Details of the accelerator head are also shown.}
\label{f:dtinternal}
\end{figure}

At the center of the DTG setup is a MF Physics Model A-211 pulsed neutron 
generator.  This neutron generator creates neutrons by
the deuterium-tritium reaction,\\
\begin{center}
\nuc{3}{H} + \nuc{2}{H} $\rightarrow$ \nuc{4}{He} + n.
\end{center}
This reaction yields isotropically distributed neutrons 
with an energy of 14.2 MeV.
The neutron generator consists of three main components,
(1) an accelerator control unit, where high voltage ($\sim$500 V),
operational interlock, and fire controls are located, (2) a pulse-forming
electronics unit, where the $\sim$100 kV pulses needed
by the accelerator are created, and (3) the accelerator head,
containing the deuterium/tritium ion source and target.
For operation at SK, where the separation between
the accelerator control unit and the accelerator head
could be up to 50 meters, the pulse-forming
electronics were repackaged and attached directly
to the accelerator head.  The combined accelerator head and pulse
electronics are encased in a stainless steel water-tight
housing (Figure~\ref{f:dtinternal}).  The housing measures
150 cm in length and 16.5 cm in diameter, fitting easily
into the calibration ports at SK.  The pulse-forming 
electronics are attached to the accelerator control unit
by a cable bundle, containing 10 coaxial cables for high voltage
and generator control signals.  These cables pass through an epoxy 
filled cable feed-through on top of a PVC plate, which serves as the
lid for the stainless steel housing.
A water leak sensor is also included in the housing
so that the generator can be removed in case of a water leak.
An ultrasonic water sensor is also attached to the top
of the steel housing and serves as an operational interlock,
preventing operation of the generator when not in the detector.
The entire DTG apparatus is moved into position in 
the detector using a custom built computer-controlled crane.

The DTG generates neutrons by colliding deuterium and tritium
ions with a fixed metal hydride target,
also containing equal parts of
deuterium and tritium.  These ions are created by a
Penning ion source, using electric and magnetic fields to
create a plasma along the source anode, and trapping 
the resulting electrons
which ionize gas in the source region (see Figure~\ref{f:dtinternal}).  
The gas pressure is regulated by the gas reservoir element.
The deuterium and tritium ions are
accelerated toward the target through an accelerating
voltage of 80-180 kV.  The target
is biased positively with respect to the accelerating anode
to prevent secondary electrons from damaging the ion source.  
The ion source, accelerating anode and target are enclosed in an evacuated
enclosure, which is in turn enclosed in a protective aluminum housing. 
The pulse-forming electronics are  enclosed in a similar 
aluminum housing and both units are filled with Shell Diala 
AX transformer oil for insulation and cooling.\footnote{The original
fluorine-based insulating fluid was replaced 
with oil to remove contamination
from \nuc{16}{N} produced inside the generator from the
(n,$\alpha$) reaction on \nuc{19}{F}.}
The neutron generator can be pulsed at a maximum rate of
100 Hz, with each pulse yielding approximately $10^{6}$ neutrons.

The 14.2 MeV neutrons produced by the DTG are energetic enough to
produce \nuc{16}{N} by the (n,p) reaction on \nuc{16}{O}\cite{mgold} in the
water of SK, which requires neutron energies greater
than $\sim$11 MeV.\cite{neutxs}.   The (n,$\alpha$)
and (n,d) reactions on \nuc{16}{O} result in the creation of stable isotopes,
while the creation of \nuc{15}{O} by the (n,2n) reaction is
energetically forbidden.  The (n,p) reaction on \nuc{17}{O}
and \nuc{18}{O} are suppressed by the low isotopic abundance
and smaller reaction cross sections, which results in yields
$<1\times 10^{-4}$ that of \nuc{16}{N}.  A simulation
of neutrons in water using GCALOR and GEANT\cite{geant} indicates
an expected \nuc{16}{N} yield, defined as the fraction of neutrons that
create \nuc{16}{N}, of 1.3\%.  The actual yield of $\sim$1\% 
found while taking data at SK agrees with this figure,
given the uncertainties in the absolute neutron
flux.  These simulations also indicate that the
mean distance these neutrons travel in 
water before they create \nuc{16}{N} is about 20 cm.

\begin{figure}
\begin{center}
\epsfig{file=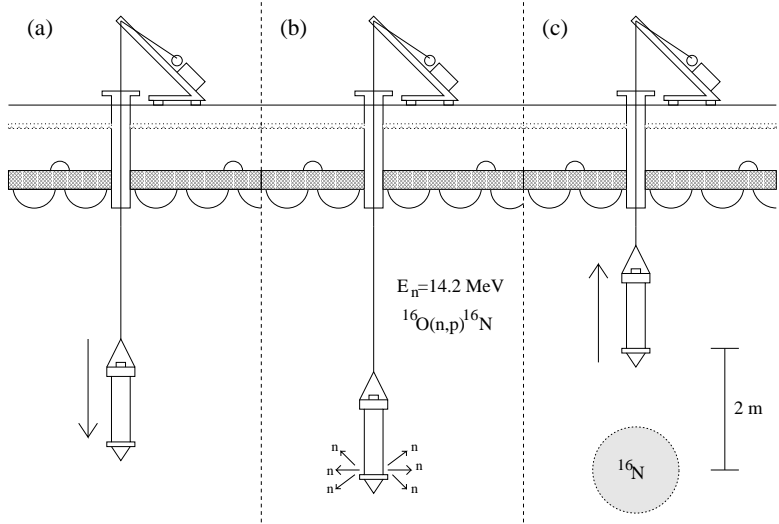,height=7cm}
\end{center}
\caption{An overview of DTG data taking.  In (a), the DTG
is lowered to the position where data is to be taken, The DTG
is fired in (b) at that location, and (c) withdrawn 2 meters 
and data collected.}
\label{f:dtoper}
\end{figure}

When taking data at SK, the DTG is lowered to a position 2 meters
above the intended \nuc{16}{N} production point, and the data taking 
cycle is started (Figure~\ref{f:dtoper}).  The data taking cycle
is controlled by computer, directing the crane, the generator
and data collection of SK.  First, the crane lowers
the DTG 2 meters, to the data collection position.  Next,
the generator is fired, creating a bubble of \nuc{16}{N} surrounding
the end of the DTG.  Every time the DTG is fired, the 
generator is pulsed 3 times, at the maximum rate of 100 Hz, 
producing $\sim$3 million neutrons. 
Third, the DTG is raised 2 meters, removing the
generator from the area containing \nuc{16}{N}.  
After the DTG is fired, $\sim$10 seconds
are required before the apparatus is completely
withdrawn, and $\sim$60\% of \nuc{16}{N} has decayed.  No
data are collected while the crane is moving to
prevent electrical noise generated by the crane from
contaminating the data. 
Once the crane has stopped moving upward, data are collected
in SK for 40 seconds.  This cycle is repeated $\sim$25 times
at a single location in the SK tank, yielding a data sample of $\sim$300,000 
\nuc{16}{N} events collected by SK.  This data sample is later analyzed
using the standard analysis tools.

\subsection{\nuc{16}{N} from the capture of $\mu^{-}$}

The creation of \nuc{16}{N} occurs naturally as a background
to the solar neutrino measurement.  A stopped $\mu^{-}$ can
be captured by a \nuc{16}{O} nucleus in the water of
the detector,
\begin{center}
\nuc{16}{O} + $\mu^{-} \rightarrow$ \nuc{16}{N} + $\nu_{\mu}$.
\end{center}
A fraction of \nuc{16}{N} created will
be in the ground state, and  beta decays with a 7.13 sec
half life.  These events are found by 
collecting events that occur in the area surrounding
the stopping point of a captured muon
and subtracting random background events.

The expected rate of \nuc{16}{N} events from muon capture per day in the inner
11.5 kton fiducial volume is given by\\
\begin{center}
$N_{ev} = N_{stopmu}(\frac{\mu^{-}}{\mu^{+}+\mu^{-}})f_{capture}f_{gs}\epsilon$,
\end{center}
where:\\
\begin{itemize}
\item
$N_{stopmu}$ is the rate of stopping muons found in the 11.5 kton
volume per day.  This rate is measured using the same stopping
muon fitter that is used in the search for \nuc{16}{N} events
and is found to be $2530 \pm 60 (stat.+sys.)$ per day.  
\item
$(\frac{\mu^{-}}{\mu^{+}+\mu^{-}})$ is the fraction of events that
are $\mu^{-}$, taken to be $0.44 \pm 0.01$\cite{mupaper}.  
\item
$f_{capture}$ is the fraction of $\mu^{-}$ that are captured on \nuc{16}{O}
before decaying, and is determined by the known capture and
decay rates\cite{pdb,suz_cap} to be $18.39\% \pm 0.01\%$.
\item
$f_{gs}$ is the fraction of captures that results in the formation
of \nuc{16}{N} in the ground state, where it will beta decay.  $f_{gs}$
is determined by the ratio of the partial capture rates
to the ground state producing levels\cite{morita,guichon}
to the total capture rate and is found to be $9.0\% \pm 0.7\%$.
\item
$\epsilon$ is the triggering and reconstruction efficiency
for \nuc{16}{N} and is determined by MC simulation to be
$64.6\% \pm 0.2\%$.  
\end{itemize}
This results in a predicted \nuc{16}{N}
event rate in the 11.5 kton fiducial volume of $11.9 \pm 1.0$ events
per day.  The inner 11.5 kton fiducial volume is chosen for
the rate analysis to avoid contamination of the stopping muon sample
from misidentification of through going muons near the
walls of the detector.  This contamination has no effect
on the solar neutrino analysis.

\subsection{Modeling the decay of \nuc{16}{N}}

Careful modeling of the \nuc{16}{N} beta decay is crucial
in order to perform accurate MC simulations of \nuc{16}{N} data. 
All decay lines with a probability of $10^{-8}$ or greater
are included.  Table~\ref{t:model} contains information about
the included decay lines.
In order to properly model the beta spectrum of
these transitions, many corrections were applied\cite{betacorr}.
These included corrections for nuclear recoil, the nuclear Coulomb field,
finite nuclear size corrections, Dirac wave function corrections, radiative
corrections and atomic screening.  Additionally, the unique first forbidden 
transitions require corrections for the 
electron spectral shape\cite{morita}.  
These corrections tend to be significant only
for very low energy electrons.  In our case, individual corrections
to the peak energy are all $<$0.2\%, and since they 
have different signs, the overall shift in the peak energy is only 0.14\%.
The electron and
gamma energies obtained from this decay simulation are then input 
into the standard SK detector simulation used in the solar neutrino
analysis.

\begin{table}
  \begin{tabular}{|c|c|c|c|c|}  \hline
    Fraction &J$^{p}_{i}\rightarrow$J$^{p}_{f}$  & $\Delta$I & Gamma Energy (MeV)
& Type\\ \hline
    66.2\%   &   $2^{-}\rightarrow3^{-}$    & +1   & 6.129 & GT allowed\\
    28.0\%   &   $2^{-}\rightarrow0^{+}$    & -2   & none & GT 1st forbidden\\
    4.8\%   &   $2^{-}\rightarrow1^{-}$    & +1   & 7.116 & GT allowed\\
    1.06\%   &   $2^{-}\rightarrow2^{-}$    & +0   & 8.872 & F+GT allowed\\
    0.012\%   &   $2^{-}\rightarrow0^{+}$    & -2   & 6.049 & GT 1st forbidden\\
    0.0012\%   &   $2^{-}\rightarrow1^{-}$    & +1   & 9.585 & GT allowed\\
	\hline
  \end{tabular}
  \caption{Summary of information used in modeling
the beta decay of \nuc{16}{N}\cite{toi}.  Gamma energies are
also included.  GT denote Gamow-Teller transitions and F denote Fermi
transitions. }
  \label{t:model}
\end{table}

\section{Results from the DTG calibration}

The DTG was installed at SK in March, 1999.  The first few
months of data taking were dedicated to engineering runs
to optimize the data taking system and
gain additional understanding of this new calibration
data.  Good, high statistics calibration data were obtained
starting in July, 1999.  At this time, a complete survey of the
detector volume with the DTG was performed. Since that time, DTG 
calibration data have been collected monthly
to monitor the long term stability of the energy scale.
Results presented here are from the July 1999 detector DTG survey.

\begin{figure}
\begin{center}
\epsfig{file=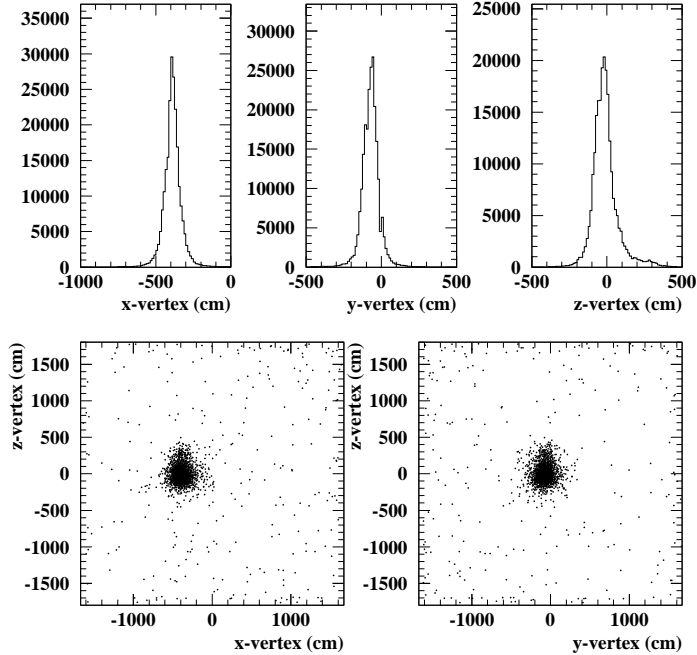,height=10cm}
\end{center}
\caption{Reconstructed vertex distributions for a typical data taking
run.  The x, y, and z-vertex position are in SK coordinates.  SK uses a
Cartesian coordinate system with the origin located at the center of
the tank.  This data
was taken at (-388.9 cm,-70.7cm ,0cm).}
\label{f:vertex}
\end{figure}

Data from the DTG are reconstructed using the same analysis tools
as the solar neutrino analysis (see Reference~\cite{skflux}).
No background subtraction is required when analyzing DTG data, as
the natural background level contributes $\ll 0.1\%$ to the data
sample.
The reconstructed vertex distributions for a typical data taking
run are presented in Figure~\ref{f:vertex}.  Histograms of
the x, y, and z vertices in SK coordinates are presented as well
as x-z and y-z scatter plots.  This data was taken at a 
nominal position of (-388.9 cm, -70.7 cm, 0 cm).  The x and y vertex
position are established by the position of the calibration port
on the SK detector, and are accurate to $\sim$3 cm.  The z vertex
position is determined by the location of the crane, and since no effort
was made during data taking to obtain the precise depth of
the DTG, the z vertex position is accurate to $\sim$25 cm.  The
vertex distributions shown in Figure~\ref{f:vertex} have fitted
peak positions from a Gaussian fit of (-389.2 cm, -72.8 cm, -27.8 cm).
A small amount of smearing can be 
seen in the upper half of the z-vertex distribution 
resulting from water displaced by the withdrawal of the DTG.
Given the spatial extent of the \nuc{16}{N} created by
the DTG, it is not the optimal tool for studies of vertex shifts
or vertex resolution.  For these studies, data from the LINAC
is used.  Reconstructed direction distributions 
are consistent with an isotropic source.

\begin{figure}
\begin{center}
\epsfig{file=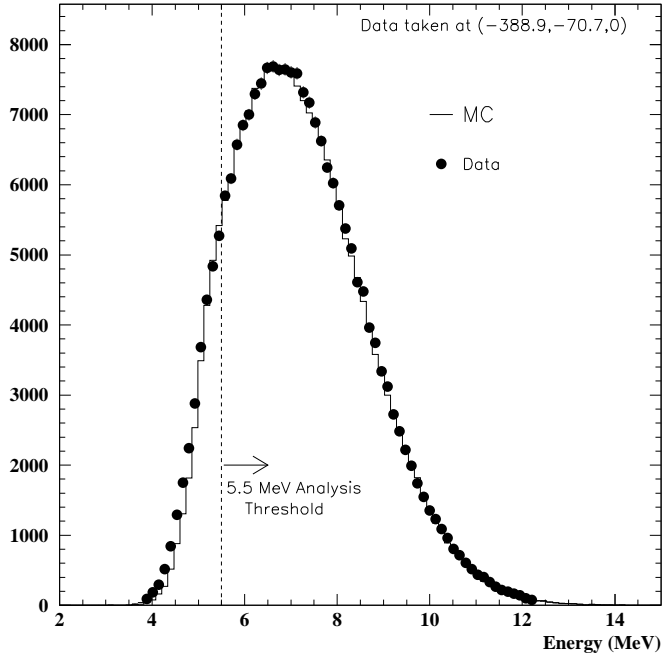,height=10cm}
\end{center}
\caption{Energy spectrum for Data and MC from a typical data taking
run at a single point in the SK tank.  
The data (points) and MC (line) are fit with a Gaussian
function only above the 5.5 MeV analysis threshold.}
\label{f:sampspec}
\end{figure}

Monte Carlo data are generated at each position where data were taken.
Reflection of Cherenkov photons from the DTG was included in the
simulation, although geometric shadowing is expected for only 0.1\%
of photons.
The data and MC events for a given position are subjected to the
same reconstruction, and a histogram of the reconstructed
energy is made for each.  The energy spectrum from a typical data
taking run is presented in Figure~\ref{f:sampspec},
along with the corresponding MC simulation.  The peak of the energy
distribution is dominated by events with a 6.1 MeV gamma ray
in coincidence with an electron with a 4.3 MeV endpoint energy.
28\% of the events contain an electron with an
endpoint energy of 10.4 MeV and are the primary source of the observed 
high energy tail.  The shape of the energy spectrum at low
energies ($<$5MeV) is primarily determined by the trigger threshold of
the detector.
These distributions are fit with a Gaussian function
between 5.5 MeV and 9.0 MeV, and the peak
positions are found.  These numbers serve as the measure
of the absolute energy scale by calculating the 
deviation of peak positions, 
$\frac{MC-DATA}{DATA}$, and this procedure is repeated
at each data position in the SK tank.

\begin{figure}
\begin{center}
\epsfig{file=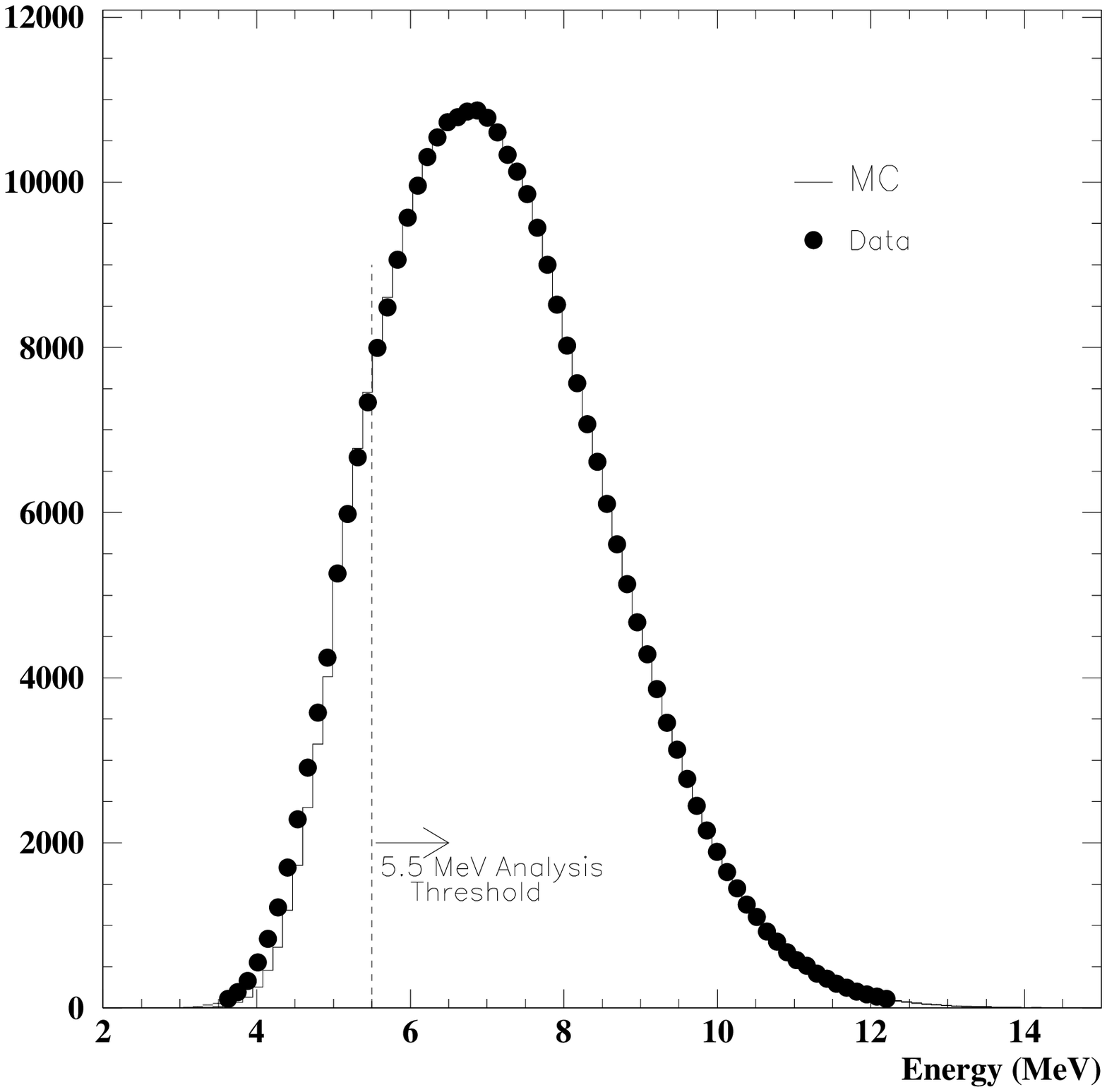,height=10cm}
\end{center}
\caption{The position-weighted average energy spectrum for
data (points) and MC(line).}
\label{f:combspec}
\end{figure}

In order to obtain a global result from the deviation measured at various
positions in the detector, a position-weighted average is
performed on the results.  Each position where DTG data are taken
is given a weight based on the geometrical fraction the volume surrounding 
that point contributes to the total 22.5 kton fiducial volume.
These weights vary from
1\%-7\% depending on the location in the detector.  Since solar neutrinos
interact uniformly throughout the entire detector volume, a 
position-weighted average provides a more realistic representation
of the detector than a simple average.
Figure~\ref{f:combspec} presents the
position-weighted average energy spectrum for data and MC. 
There is excellent agreement between the two.

\begin{figure}
\begin{center}
\epsfig{file=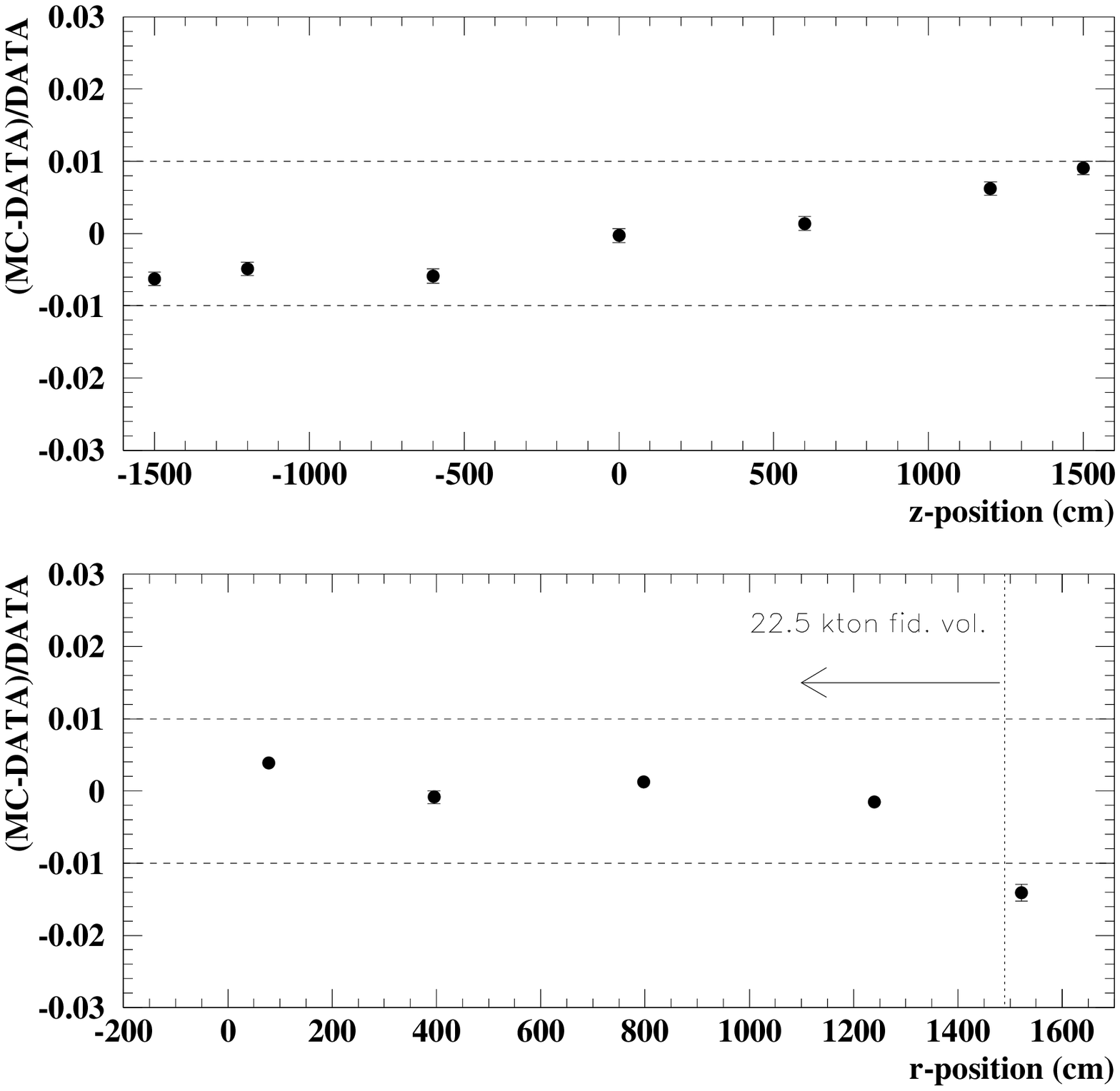,height=10cm}
\end{center}
\caption{Position dependence of the energy scale from DTG data.
At each r and z vertex position, a position-weighted average
over all z and r positions, respectively, is performed. Only
statistical errors are shown.  Dashed lines indicate a deviation
of $\pm$1\%.}
\label{f:energypos}
\end{figure}

\begin{figure}
\begin{center}
\epsfig{file=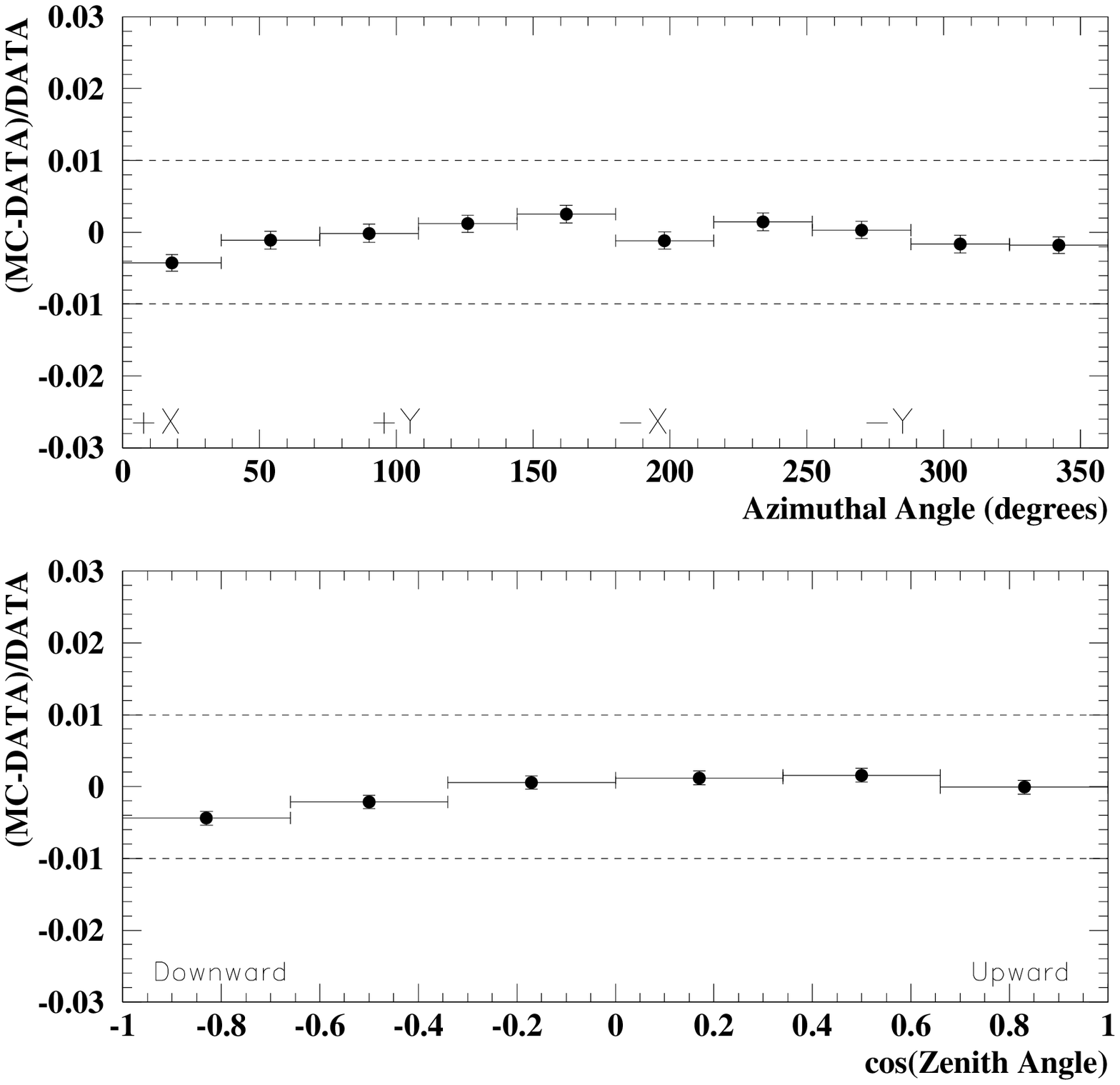,height=10cm}
\end{center}
\caption{Angular dependence of energy scale from DTG data obtained
from a position-weighted average over all positions in the fiducial volume.
An azimuthal angle of 0 degrees corresponds to the +X axis of
SK.  Only statistical errors are shown.  Dashed lines indicate a deviation
of $\pm$1\%.}
\label{f:energyang}
\end{figure}

The data from the DTG are also used to study the position and direction
dependence of the energy scale.  DTG data was taken in 6 different
calibration port locations, at 7 depths per port, providing
a large sampling of the detector volume.  The position dependence
of the energy scale, shown in Figure~\ref{f:energypos},
is presented as a function of radial distance (r) and height (z)
in the detector, by performing a position-weighted average
over z and r, respectively.
No significant variation in r is seen over the
fiducial volume.  While there is a slight
variation with height in the detector, it is within
the systematic errors for the energy scale.
The direction dependence of the energy scale
is obtained by dividing the data in subsets at each position
based on the reconstructed direction,
performing a Gaussian fit on the energy spectrum of each
subset, and then performing
a position-weighted average over all DTG data positions in the
detector for each direction.  The direction dependence is studied as a function
of zenith angle, measured with respect to the vertical (z) axis
of the detector, and as a function of azimuthal angle, measured
in the x-y plane.
The resulting  angular dependence of the energy scale is
presented in Figure~\ref{f:energyang}.  In both
cases, the variation in the energy scale in direction
within the fiducial volume is less than 1\%.

\begin{figure}
\begin{center}
\epsfig{file=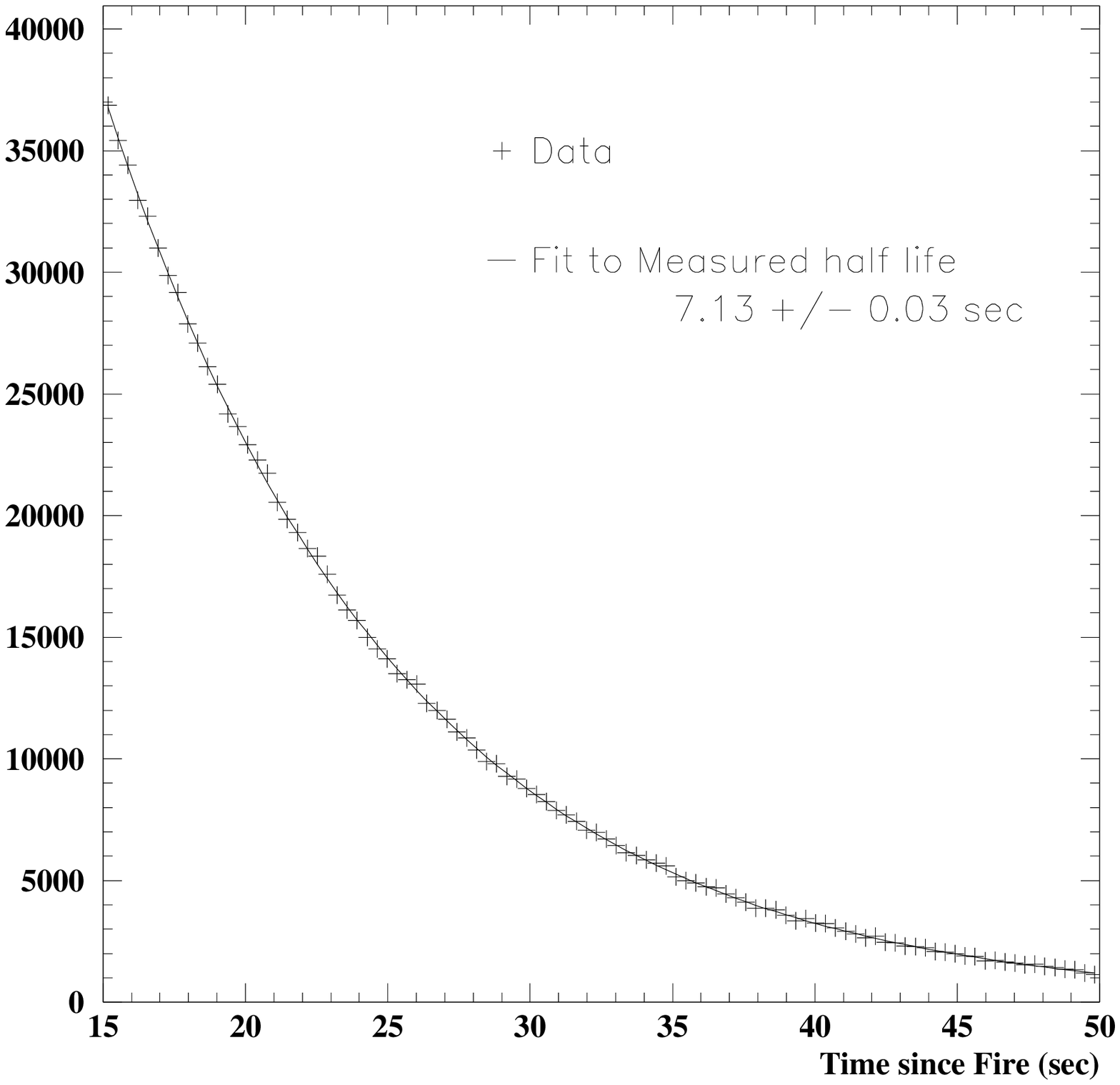,height=10cm}
\end{center}
\caption{Distribution of time since generator fire for
several DTG runs.  The data (crosses) and best fit (line)
are shown.  The expected half life is $7.13 \pm 0.02$ sec.
Data with time since fire less than 15 seconds are not shown
to ensure all data is taken after the crane has been fully withdrawn.}
\label{f:halflife}
\end{figure}

As a check for background contamination from other
nuclides, the half life of \nuc{16}{N} is measured using the
collected data.  The time since generator fire for each
event collected is plotted, and the data from several 
positions are combined for additional statistical weight.
The histogram of this decay time and the calculated best
fit are shown in Figure~\ref{f:halflife}.  The best fit 
half life of $7.13 \pm 0.03$ sec is in excellent agreement
with the expected value\cite{toi} of $7.13 \pm 0.02$ sec,
indicating a clean sample of \nuc{16}{N} is obtained.

The systematic errors for the DTG calibration for the 
absolute energy scale measurement are 
summarized in Table~\ref{t:syser}.  The systematic error
for shadowing of Cherenkov photons 
by DTG related equipment after it is withdrawn
is determined by the fraction of photons 
that could be absorbed and is conservatively 
estimated to be $\pm0.1\%$.  A simulation of neutrons
in the DTG setup indicates that small amounts of
background isotopes are created,
including \nuc{24}{Na}, \nuc{62}{Co}, and \nuc{28}{Al}.
Most nuclides have long half lives and/or insufficient
energy to trigger SK, but a MC simulation indicates
a small amount of gamma contamination is possible, and a
systematic error of $\pm0.1\%$ is conservatively chosen.
The DTG data selection systematic error results from  
a vertex position cut made to the data 
to remove background events occurring near the walls
of the detector from the data sample.

The total position averaged 
energy scale deviation, $(\frac{MC-DATA}{DATA})$,
measured during the July 1999 survey of the detector
is found to be $-0.04\%\pm0.04\%(stat.)\pm0.2\%(sys.)$,
indicating excellent overall agreement of the DTG data with the 
LINAC-based MC simulation.

\begin{table}
\begin{tabular}{|l|l|}  \hline
Contamination from natural background&$<0.01\%$\\
\nuc{16}{N} MC decay modeling&$\pm0.1\%$\\
Unmodeled decay lines&$<0.01\%$\\
Shadowing of Cherenkov photons&$\pm0.1\%$\\
DTG data selection systematic&$\pm0.1\%$\\
DTG related radioactive background&$\pm0.05\%$\\
\hline
Total Systematic Error&$\pm0.2\%$\\
\hline
\end{tabular}
\caption{Summary of systematic errors from the DTG calibration.}
\label{t:syser}
\end{table}

\section{Results from the $\mu^{-}$ Capture \nuc{16}{N} Analysis}

\nuc{16}{N} data resulting from the capture of stopped
$\mu^{-}$ are simultaneously collected during normal data
collection runs with solar neutrino events.  In order to extract
these few events per day from the data set, 
a specialized data search is implemented.
First, stopping muon events are found and the stopping positions
reconstructed.  These fit results
are used as an input, along with the input data for the
solar neutrino analysis, to the \nuc{16}{N} search program.
The search program finds stopping muon events that
are not followed by a decay event in 100$\mu$s, 
as a muon that decays can not be captured on \nuc{16}{O},
preventing additional background contributions.  Once an
``undecayed'' muon is found,
any events in the solar neutrino data sample are saved that occur
within 335 cm of the stopping muon point, as well as in
a time window of 100 ms to 30 seconds following the stopping
muon.  The distance of 335 cm
is used as it is large enough to contain all expected
signal events while keeping contributions from random backgrounds
as small as possible.
No limit is placed on the number of candidate signal
events a stopping muon can produce to prevent biasing from
the random background in SK.  The result of this search
is called the signal sample, and contains the \nuc{16}{N}
events as well as natural background events.  
To account for this natural background, a so-called background sample
is also obtained by offsetting the times of the
``undecayed'' muons by 100 seconds into the future and again
performing the same search.  

Once the signal and background samples are compiled, the solar neutrino
analysis algorithms are applied to all events.  The effects 
of the natural background are removed by performing a 
statistical subtraction of the results
for the background sample from the signal sample.  
In 1003.8 days of data,
41,059 signal sample events and 20,025 background sample events are found,
resulting in an excess of 21,034 events attributed to \nuc{16}{N} 
decay.  In the inner 11.5 kton
fiducial volume, 17,714 signal, with 6267 background events
are found in this same live-time period,
resulting in a  rate of \nuc{16}{N} events 
of 11.4$\pm$0.2 events/day.  This result agrees
well with the predicted rate of 11.9$\pm$1.0 events/day and 
indicates that no loss of signal is found for events in
the same energy range of solar neutrinos.

\begin{figure}
\begin{center}
\epsfig{file=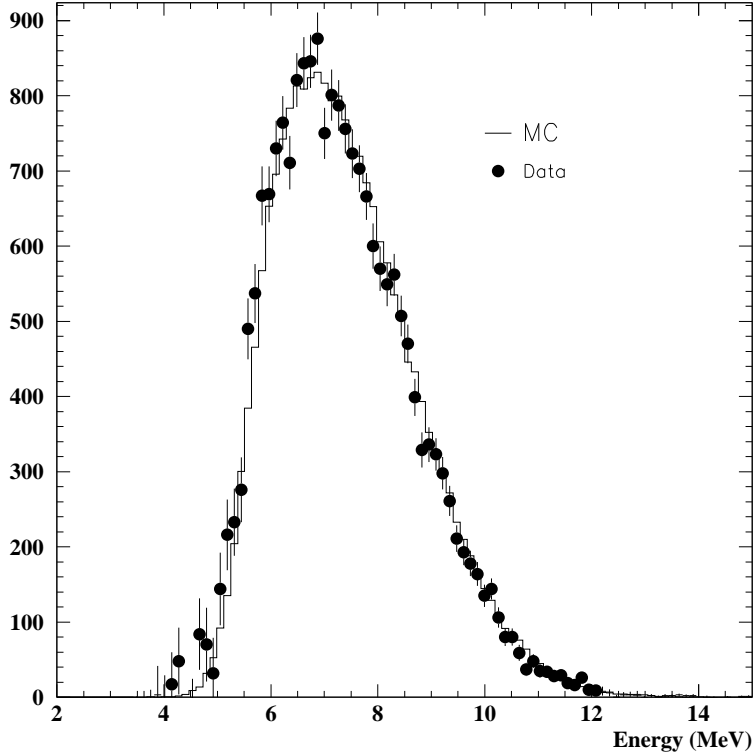,height=10cm}
\end{center}
\caption{\nuc{16}{N} background subtracted energy distribution(points)
along with MC expectations(line).}
\label{f:nat_energy}
\end{figure}

These data are also used to check the absolute energy scale
of the detector by comparing the energy distribution
of the background subtracted sample to MC.
A MC event sample is generated with measured
decay electron positions used as input vertices.  These
events are reconstructed using the same tools
as the signal and background samples.
The background subtracted energy distribution is 
presented in Figure~\ref{f:nat_energy} along with
the results from the MC. The fitted peak position
for data and MC are found and compared, yielding a 
deviation $(\frac{MC-DATA}{DATA})$ of +0.9\%$\pm$0.7\%.  

\begin{figure}
\begin{center}
\epsfig{file=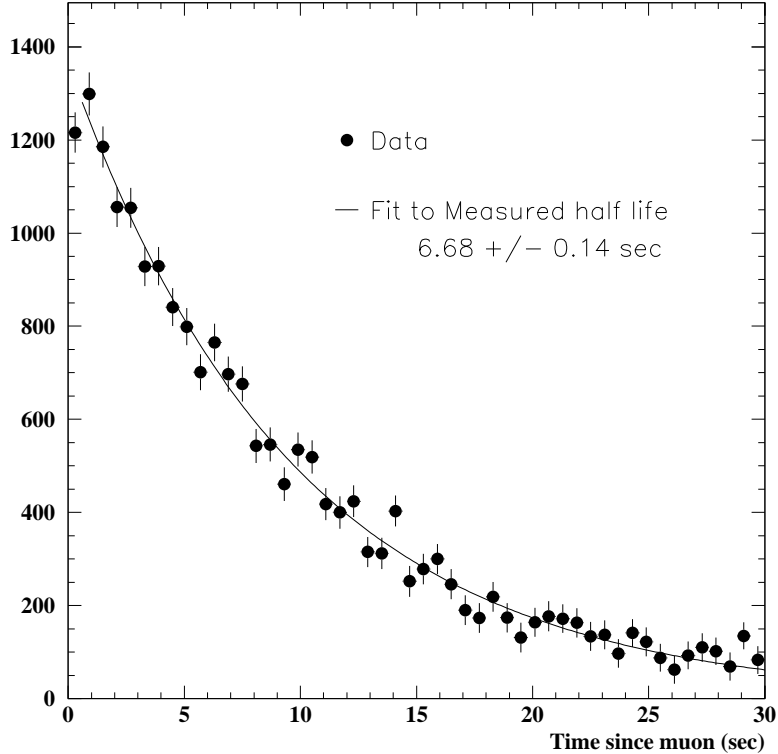,height=10cm}
\end{center}
\caption{Background subtracted distribution of time since
stopping muon for \nuc{16}{N} candidate events(points) and best
fit to data(line).}
\label{f:nat_halflife}
\end{figure}

The background subtracted distribution of time since
the stopping muon for \nuc{16}{N} events
is used to measure the half life, and is presented in 
Figure~\ref{f:nat_halflife}.  The measured half life
of $6.68\pm0.14$ sec is in poor agreement with the expected
half life of $7.13\pm0.02$ sec, with a 3.2 sigma difference.  
The poor agreement of the half life 
seems to indicate the presence of
contamination in the data sample.  \nuc{15}{C} has
been suggested, with a half life of 2.45 sec.
It can be created by proton emission of some excited 
states of \nuc{16}{N} formed during the capture of the $\mu^{-}$.
Little is known about the highly excited states of the
\nuc{16}{N} nucleus\cite{nucref}, and this is taken as a systematic
error for this analysis.  The deviation of the measured half 
life from the accepted value can be interpreted as 
a $\sim$4\% contamination from \nuc{15}{C} and
a MC of \nuc{15}{C} decay events indicates that this would shift
the energy distribution by -0.4\%.
This type of contamination is not energetically possible
for the DTG data sample.  Additionally, the systematic error
in modeling the decay of \nuc{16}{N}
from the DTG analysis also applies here, resulting in a total
energy scale deviation for the $\mu^{-}$ capture \nuc{16}{N}
of $+0.9\%\pm0.7\%(stat.)^{+0.1\%}_{-0.5\%}(sys.)$

\section{Conclusions}

The absolute energy scale of the Super-Kamiokande detector
for solar neutrinos has been verified by the decay of \nuc{16}{N}.
In particular, \nuc{16}{N} resulting from the (n,p) reaction
of 14.2 MeV neutrons on water provides a high statistics data sample
with comparable systematic uncertainties to the LINAC.  Our
analysis shows
excellent agreement (better than $\pm1\%$) between data 
and a LINAC-tuned MC simulation.  
The position and direction dependence of the energy
scale have also been studied and are within our systematic
uncertainties for the energy scale of the detector.
The analysis of \nuc{16}{N} events resulting from the
capture of stopped $\mu^{-}$ shows good agreement
between the expected and measured rates of events and verifies
solar neutrino flux measurements.

DTG data are periodically being collected to study the time dependence
of the energy scale, and the DTG will continue to be used as a 
calibration tool for Super-Kamiokande.  Future calibrations that
combine the LINAC and DTG could result in smaller systematic
uncertainties in the energy scale and more accurate measurements
of the solar neutrino energy spectrum.

\section{Acknowledgment}  \label{secAK}

We gratefully acknowledge the cooperation of the Kamioka Mining and
Smelting Company.
This work was partly supported by the Japanese Ministry of Education,
Science and Culture, the U.S. Department of Energy and the 
U.S. National Science Foundation.

\end{document}